\documentclass[12pt]{article}
\usepackage{setspace}
\usepackage{amsmath}
\usepackage{graphicx}
\usepackage{cases}
\usepackage{psfrag}
\usepackage{amssymb}
\usepackage{color,graphics,epsfig}
\usepackage{appendix}
\usepackage{fancyhdr}
\usepackage{textcomp}
\usepackage{float}
\usepackage{caption}
\usepackage{subcaption}
\usepackage[square,sort,comma,numbers]{natbib}

\def\be{\begin{equation}}
\def\ee{\end{equation}}
\def\bea{\begin{eqnarray}}
\def\eea{\end{eqnarray}}

\begin{document}

\title{Evolutionary dynamics and eigenspectrum of confluent Heun equation}

\author{Kavita Jain and Archana Devi\\
Theoretical Sciences Unit, \\
Jawaharlal Nehru Centre for Advanced Scientific Research,\\
 Bangalore 560064, India}

 \maketitle

\clearpage

\abstract{
We consider a biological population evolving under the joint action of selection, mutation and random genetic drift. The evolutionary dynamics are described by a one-dimensional Fokker-Planck equation whose eigenfunctions obey a confluent Heun equation. These eigenfunctions are expanded in an infinite series of orthogonal Jacobi polynomials and the expansion coefficients are found to obey a three-term recursion equation. Using scaling ideas, we obtain an expression for the expansion coefficients and an analytical estimate of the number of terms required in the series for an accurate determination of the eigenfunction. 
The eigenvalue spectrum is studied using a perturbation theory for weak selection and numerically for strong selection. In the latter case, we find that the eigenvalue for the first excited state exhibits a sharp transition: for mutation rate below one, the eigenvalue increases linearly with increasing mutation rate and then remains a constant; higher eigenvalues are found to display a more complex behavior. }

\vspace{2pc}

\maketitle

\clearpage

\section{Introduction}
\label{intro}

Biological evolution has shaped the genetic diversity that we see today on earth \cite{Crow:1970,Charlesworth:2010}. The basic processes that drive evolution include selection, mutation and random genetic drift. While selection decreases the genetic diversity, random mutations increase it. Besides these deterministic processes, stochasticity arising, for example, due to finite carrying capacity 
plays an important role in determining the evolutionary fate of a population. With growing interest in analyzing the time series data obtained from experiments or field studies \cite{Gutenkunst:2009,Taus:2017}, and addressing fundamental questions about evolution in changing environments \cite{Devi:2020}, it is important to understand the evolutionary dynamics of a population under the joint action of mutation, selection and genetic drift. 

In the absence of selection, the dynamics are completely understood \cite{Kimura:1964}, as the Fokker-Planck equation for the distribution of allele frequency  obeys the Gauss hypergeometric equation which has the nice property that its Frobenius series expansions lead to two-term recurrence relations for the expansion coefficients and for which the connection formulae that connect its two local solutions are known \cite{Erdelyi:1953a}. But even in the simplest setting where only a single locus in a genome is under selection, the complete evolutionary dynamics are not explicitly known. 

In this article, we study the allele frequency distribution in a finite population in which a single, diallelic locus is under time-independent selection and mutations between the two alleles occur at equal rates. To the best of our knowledge, it has not been previously recognized that the eigenfunctions of the Fokker-Planck equation for the frequency distribution obey  the {\it confluent Heun equation} \cite{heunsite,Hortacsu:2018}. 
This equation has appeared in diverse physical contexts to describe the eigenspectrum of hydrogen molecule ion \cite{Jaffe:1934} and quantum Rabi model \cite{Xie:2017}, quasi-normal modes of black holes \cite{Leaver:1985,Fiziev:2011}, relaxation dynamics of a polymer \cite{Vincenzi:2006}, etc. Heun equation and its confluent forms are an area of active interest in mathematics community also as the Heun functions generalize the well studied hypergeometric functions.  The progress in understanding Heun functions has, however, been slow as Frobenius series or other orthogonal polynomial expansions of this class of equations lead to {three-term} recurrence relations for the expansion coefficients, and the connection formulae for the local solutions are not explicitly known \cite{Lay:1998}. 

For these reasons, here we study the eigenspectrum of the confluent Heun operator of interest, mainly, numerically. 
As the stationary state corresponding to the zero eigenvalue is exactly known \cite{Kimura:1964}, we will focus on the excited states and study them using an orthogonal polynomial expansion of the eigenfunctions. Such an expansion has previously been carried out in \cite{Song:2012}; however, these authors obtained a five-term recurrence relation for the expansion coefficients and did not provide any insights into the solution. Here, we obtain a three-term recurrence equation and find that for strong selection, there is a {\it dynamic transition} as the relaxation time - which is inversely proportional to the eigenvalue for the first excited state - initially decreases with mutation rate and then becomes a constant at a finite mutation rate. We also show that the expansion coefficients have a scaling form which allows us to estimate the number of terms that contribute significantly to the orthogonal series for the eigenfunctions.  

\section{Model}
\label{model}

We consider a haploid population of $N$ individuals that evolves in time according to the standard Moran process \cite{Ewens:2004}. In  generation $t$, an individual is chosen to give birth with a probability equal to its fitness relative to that of the population while any individual (including the parent) can die with an equal probability. Thus, if the wild type individual has a fitness $1$ and the mutant's fitness is $1+s$, the mutant produces a copy of itself with a probability $(1+s) i/{\bar w}$ when $i$ mutants are present in the population at time $t$, where ${\bar w}(t)=(1+s)i+(N-i)$ is the average fitness of the population; similarly, the wild type individual replicates with a probability $(N-i)/{\bar w}$. After selection and reproduction, an offspring mutates to the other type with a probability $u$. Thus, the population evolves under the joint action of selection, mutation and random genetic drift.

We are interested in understanding the dynamics of the probability distribution of the mutant allele frequency, $p=i/N$ in the scaling limit of weak selection, weak mutation ($s, u \to 0$) and large population size ($N \to \infty$) with finite $N s, N u$. Using the mean and the variance of the change in the mutant frequency in the above scaling limits, one can show that the frequency distribution $\psi(p,\tau)$ obeys the following forward Fokker-Planck equation \cite{Kimura:1964,Risken:1996}, 
\be
\frac{\partial  \psi}{\partial \tau} = - \frac{\partial J}{\partial p}=- \frac{\partial }{\partial p} \left[\sigma p q \psi + \mu (q-p) \psi - \frac{\partial }{\partial p} \left(p q \psi\right)\right], 
\label{ufullFP}
\ee
where $q=1-p$ is the wild type frequency, $J(p,\tau)$ is the probability current, and $\tau=t/(2 N), \mu =  2 N u ,\sigma = 2 N s$ are the scaled time, mutation rate and selection rate, respectively. Below we will study the dynamics of the allele frequency distribution  subject to the reflecting boundary conditions, 
\be
J(0,\tau)=J(1,\tau)=0, 
\label{bdry}
\ee
at all times. 

Some comments are in order: in the following, we will assume that $s \geq 0$ since for $s < 0$, equation (\ref{ufullFP}) is obeyed by the allele frequency $q$. While the Moran process described above assumes overlapping generations, the Wright-Fisher process for non-overlapping generation also obeys (\ref{ufullFP}) when $N$ is replaced by $2 N$. Instead of the forward equation (\ref{ufullFP}), one may study the corresponding backward equation \cite{Song:2012}; however, the eigenspectrum of both equations is same \cite{Risken:1996}. For these reasons, it is sufficient to focus on the above forward equation (\ref{ufullFP}).

We first write $\psi (p,\tau)= \sum_{\ell=0}^\infty A_\ell  e^{-\lambda_\ell \tau}  \phi_\ell({p})$ where the coefficient $A_\ell$ is determined by the initial condition $\psi(p,0)$. From (\ref{ufullFP}), we find that the time-independent eigenfunction $\phi_\ell(p)$ obeys the following eigenvalue equation,
\be
-\left(\sigma p q \phi_\ell+ \mu (q-p)\phi_\ell \right)' +\left(p q \phi_\ell \right)''=-\lambda_\ell  \phi_\ell, 
\label{eigenval}
\ee
where a prime denotes a derivative with respect to $p$. 
As the differential operator in (\ref{eigenval}) is of Sturm-Liouville form \cite{Mathews:1970}, we are guaranteed to have real eigenvalues and a complete set of orthonormal eigenfunctions for which 
\be
\int_0^1 dp \rho(p) \phi_{\ell}(p)  \phi_{\ell'}(p) \propto \delta_{\ell,\ell'},
\label{orthonorm}
\ee
where the weight function, $\rho(p)\propto \phi_0^{-1}$ (see below). 

 In equilibrium where the probability current $J$ vanishes for all $p$, the probability distribution is given by \cite{Wright:1937,Kimura:1964}
\be 
\phi_0(p)  \propto  (p q)^{\mu-1} e^{\sigma p}, 
\label{equil}
\ee
where the proportionality constant determined by (\ref{orthonorm}) reduces to the normalization condition $\int_0^1 dp \phi_0(p)=1$.   For $\sigma=0$, the above equilibrium distribution is U-shaped for $\mu < 1$ and bell-shaped for $\mu > 1$; however, for large $\sigma$, the stationary distribution $\phi_0 (p)$ for the favored allele peaks close to $p=1$. 


\section{Confluent Heun equation for the eigenfunctions}

We will now focus on (\ref{eigenval}) which can be rewritten in the standard form as 
\be
\phi_\ell''(p)+\left(\frac{2-\mu}{p-1}+\frac{2-\mu}{p}- \sigma \right) \phi_\ell'(p)+ \left(\frac{\nu_\ell}{p}+\frac{-2 \sigma- \nu_\ell}{p-1} \right) \phi_\ell(p)=0, 
\label{heun}
\ee
where $\nu_\ell = 2 \mu-\sigma-2 +\lambda_\ell$. 
Equation (\ref{heun}) is a (singly) confluent Heun equation which has regular singular points at $p=0, 1$ and an irregular singularity of rank $1$ at infinite $p$ \cite{heunsite,Fiziev:2010,Hortacsu:2018}. 
The following special cases of (\ref{heun}) are known  \cite{Kimura:1964}: for $\sigma=0$ (no selection), (\ref{heun}) reduces to the Gauss hypergeometric equation which has regular singular points at zero, one and infinity, and for $\sigma=\mu=0$ (only genetic drift), the eigenfunction $\phi_\ell$ obeys the Gegenbauer equation. For nonzero selection but zero mutation rate, $V_\ell(z)=e^{-\sigma p/2} \phi_\ell(p)$ with $p=(1-z)/2$ obeys the oblate spheroidal  equation \cite{Chu:1941}. However, except for the stationary state (\ref{equil}) corresponding to eigenvalue zero, neither the eigenvalue spectrum nor the eigenfunctions are known for the above confluent Heun differential operator. 

The general solution of (\ref{heun}) can be expanded as a Frobenius series about $p=0$ \cite{Leaver:1986} which, on imposing the reflecting boundary condition (\ref{bdry}) at $p=0$ gives $\phi_\ell(p)=p^{\mu-1} F_\ell(p)$ for $|p| < 1, {\mu > 0}$ where $F_\ell$ is an analytic function (see Appendix~\ref{app_frob}). Due to the self-adjoint nature of the Heun operator, we can expand $F_\ell$ as a linear combination of suitable orthogonal functions \cite{Mathews:1970}; a power series expansion of $F_\ell$ is given in Appendix~\ref{app_frob}. 

An orthogonal expansion of (\ref{heun}) with boundary conditions (\ref{bdry}) has been carried out in \cite{Song:2012} by writing the distribution function, $\phi_\ell(p)=\sum_{n} {\hat a}_n^{(\ell)} {\hat G}_n(p)$ where ${\hat G}_n=(p q)^{\mu-1} e^{\sigma p/2}  P_n^{(\mu-1,\mu-1)}(1-2p)$ and  $P_n^{(\alpha,\beta)}(x)$ denotes Jacobi polynomial of order $n$ that obey the following orthogonality relation \cite{Abramowitz:1964}, 
\be
\int_{-1}^1 dx (1-x)^\alpha (1+x)^\beta P_n^{(\alpha,\beta)}(x) P_{n'}^{(\alpha,\beta)}(x) =h_n \delta_{n,n'}. 
\label{jacorth}
\ee
The above ${\hat G}_n$  has the property that they are orthogonal with respect to the same weight function as that for $\phi_\ell$, namely, $\rho$. But this choice leads to a $5$-term recursion relation for the coefficients ${\hat a}_n^{(\ell)}$ that are not amenable to analytical calculations.

Here, we write $\phi_\ell(p)=\sum_n a_n^{(\ell)} G_n$ with $G_n=(p q)^{\mu-1} P_n^{(\mu-1,\mu-1)}(1-2 p)$ that, as can be verified using (\ref{jacorth}), are orthogonal with respect to the weight function $(p q)^{1-\mu}$.  As detailed in Appendix~\ref{app_rec}, our choice of the orthogonal basis leads to 3-term recursion relations for $a_n^{(\ell)}$ (see also \cite{Vincenzi:2006}), and we can write 
\be
\phi_\ell(p)=(p q)^{\mu-1} \sum_{n=1}^\infty c_n^{(\ell)} \frac{\Gamma(n) }{\Gamma(n+\mu-1)} P_{n-1}^{(\mu-1,\mu-1)}(1-2 p), 
\label{jacobi}
\ee
which satisfies the boundary condition (\ref{bdry}) at $p=1$ also. The expansion coefficients $c_n^{(\ell)}$'s are determined recursively through the following equations, 
\bea
{\lambda}_\ell c_1^{(\ell)}  &=& 0 \label{c1eqn},  \\
T_-(n) c_{n-1}^{(\ell)}   + T_0(n) c_n^{(\ell)}  +   T_{+}(n) c_{n+1}^{(\ell)} &=& - \lambda_{\ell}  c_n^{(\ell)} ~,~n \geq 2, 
\label{cneqn} 
\eea
where
\bea
T_- (n)&=& \frac{ \sigma (2 \mu + n-2) ( 2 \mu + n-3)}{10 - 4 \mu - 4 n} < 0, \\
T_+(n)&=& \frac{\sigma n (1-n)}{2 - 4 \mu - 4 n} > 0, \\
T_0(n)&=& (1-n) (2 \mu+n-2) < 0. 
\eea

As already stated, the stationary state corresponding to $\lambda_0=0$ is given by (\ref{equil}) and it can be used to find the expansion coefficients $c_n^{(0)}$, as described in Appendix~\ref{app_ss}. In the following, we are interested in excited states with nonzero eigenvalues; due to (\ref{c1eqn}), this means that $c_1^{(\ell)}=0$ for $\ell=1, 2,...$. Furthermore, for large $n$, the ratio $r_n=c_{n+1}/c_n$ has following linearly independent solutions,
\be
 r_n^+ \sim\frac{4n}{\sigma}, ~~r_n^-\sim-\frac{\sigma}{4n}
 \label{rasym}
\ee
for any $\ell$. As the (minimal) solution $r_n^-$ ensures the convergence of the continued fraction method for evaluating eigenvalues \cite{Gautschi:1967} (see also Sec.~\ref{num}), we have the boundary condition that the expansion coefficients vanish at large $n$. 

In summary, (\ref{cneqn}) along with boundary conditions $c_1^{(\ell)} , c_{K+2 \to \infty}^{(\ell)} =0$  defines an eigenvalue problem, ${\bf T} \vec{c}^{(\ell)} =-\lambda_{\ell}  \vec{c}^{(\ell)}, \ell \geq 1$ for eigenvector $\vec c^{(\ell)} $ and eigenvalue $-\lambda_\ell$ where ${\bf T}$ is $K$-dimensional square matrix.


\section{Eigenvalue problem: numerical analysis}
\label{num}

For the ratio $r_n^{(\ell)}$, the recursion equation (\ref{cneqn}) can be written as 
\be
r_{n-1}^{(\ell)}=\frac{j_n}{k_n+r_n^{(\ell)}}=\frac{j_n}{k_n+\frac{j_{n+1}}{k_{n+1}+r_{n+1}^{(\ell)}}}, 
\ee
where $j_n=-T_-(n)/T_+(n)$ and $k_n^{(\ell)}=(T_0(n)+\lambda_\ell)/T_+(n)$. Continuing in this manner, we have 
\be
r_2^{(\ell)}=\cfrac{j_3}{k_3^{(\ell)}+\cfrac{j_{4}}{k_{4}^{(\ell)}+\cfrac{j_5}{k_5^{(\ell)}+...}}}=-k_2^{(\ell)}, 
\label{cont}
\ee
where we have used the boundary condition $c_1^{(\ell)}=0, \ell > 0$. Starting from $r_n^-$ in (\ref{rasym}), the above continued fraction was used to calculate $\lambda_\ell$ and $r_n^{(\ell)}$ numerically \cite{Gautschi:1967}, and we found the ratio $r_n^{(\ell)}$ to be negative for all $n$. From these ratio of expansion coefficients, the eigenfunction $\phi_{\ell}(p)$ was obtained by carrying out the sum over $K+1$ terms in (\ref{jacobi}). One can also diagonalize matrix ${\bf T}$  to find eigenvalues and eigenfunctions. 
To obtain numerical results, both these methods were applied to  finite-dimensional matrix ${\bf T}$ of large size $K$. 

\subsection{Eigenvalues}

\begin{figure}
\centering
\includegraphics[width=0.8\textwidth]{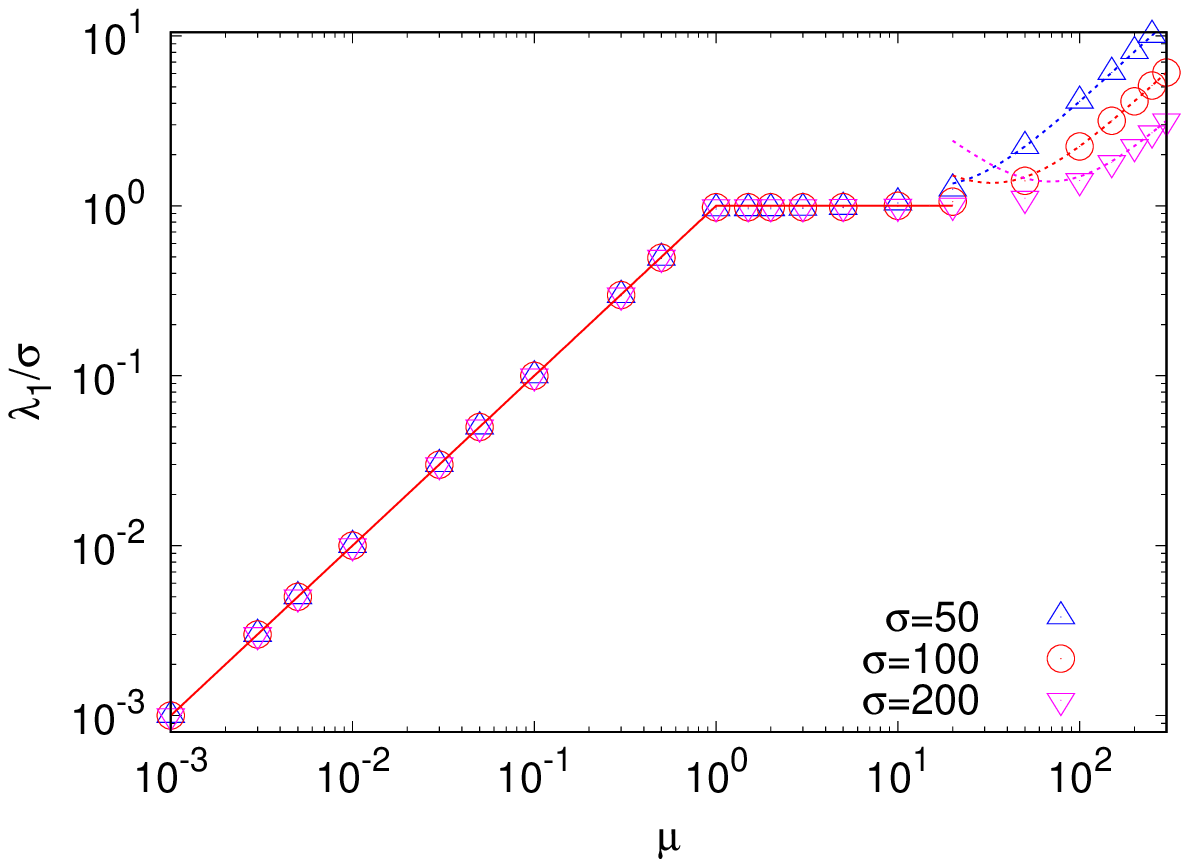}
\caption{Variation of eigenvalue $\lambda_1$ with mutation rate $\mu$ for various values of selection parameter $\sigma$ and $K=1000$. The points are obtained numerically using (\ref{cneqn}), and the solid and dotted lines show the conjecture (\ref{firstlambda}) for strong selection and analytical expression (\ref{pert1}) for weak selection, respectively.}
 \label{fig_lam1}
\end{figure}

\begin{figure}
\centering
\includegraphics[width=0.8\textwidth]{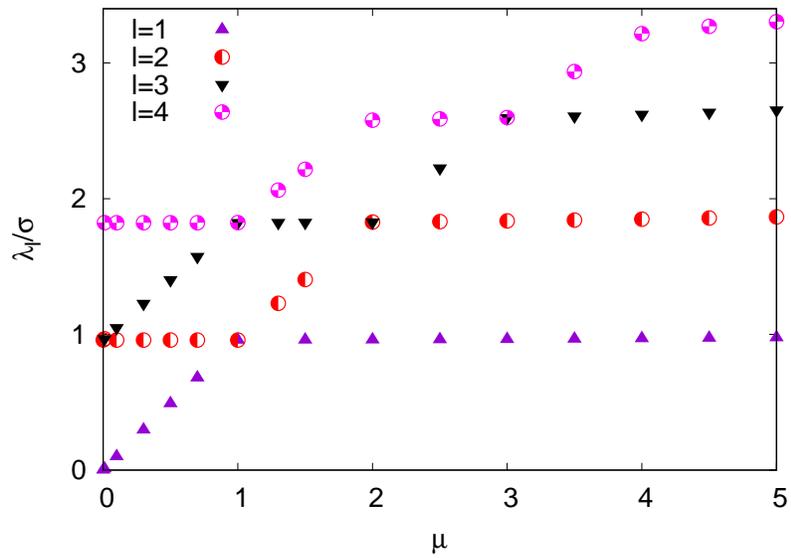}
\caption{Eigenvalue spectrum for strong selection obtained numerically using (\ref{cneqn}) for $\sigma=50$ and $K=1000$.}
\label{fig_strsel}
\end{figure}

Figure~\ref{fig_lam1} shows the numerical results for the eigenvalue $\lambda_1$ as a function of the mutation rate $\mu$ and large selection strengths. For $\mu \gg \sigma$ where selection is weak relative to mutation, the eigenvalue varies with $\mu$  essentially the same way as that in the absence of selection  since, as discussed in Sec.~\ref{weak}, the correction to $\lambda_1$ due to selection is quadratic in $\sigma/\mu$. But for $\mu \ll \sigma$, the eigenvalue increases linearly with $\sigma$ (see Sec.~\ref{strong}), and our numerical results in Fig.~\ref{fig_lam1} suggest that 
\be
\frac{\lambda_1}{\sigma} \stackrel{\sigma \to \infty}{\longrightarrow}
\begin{cases}
\mu, & \mu \leq  1\\
1, & \mu > 1
\end{cases}.
\label{firstlambda}
\ee
We have also studied the corrections to the above conjectured eigenvalue, and find that they approach the asymptotic values as  $1/\sigma$ (data not shown). The higher eigenvalues shown in Fig.~\ref{fig_strsel} exhibit a more complex behavior. The eigenvalue $\lambda_2$ remains a constant for $\mu < 1$, increases linearly for $1 < \mu < 2$ and is a constant for $\mu \geq 2$; a similar pattern is seen for $\lambda_3$ and $\lambda_4$. 
\subsection{Eigenfunctions}

\begin{figure}
  \includegraphics[width=0.8\textwidth]{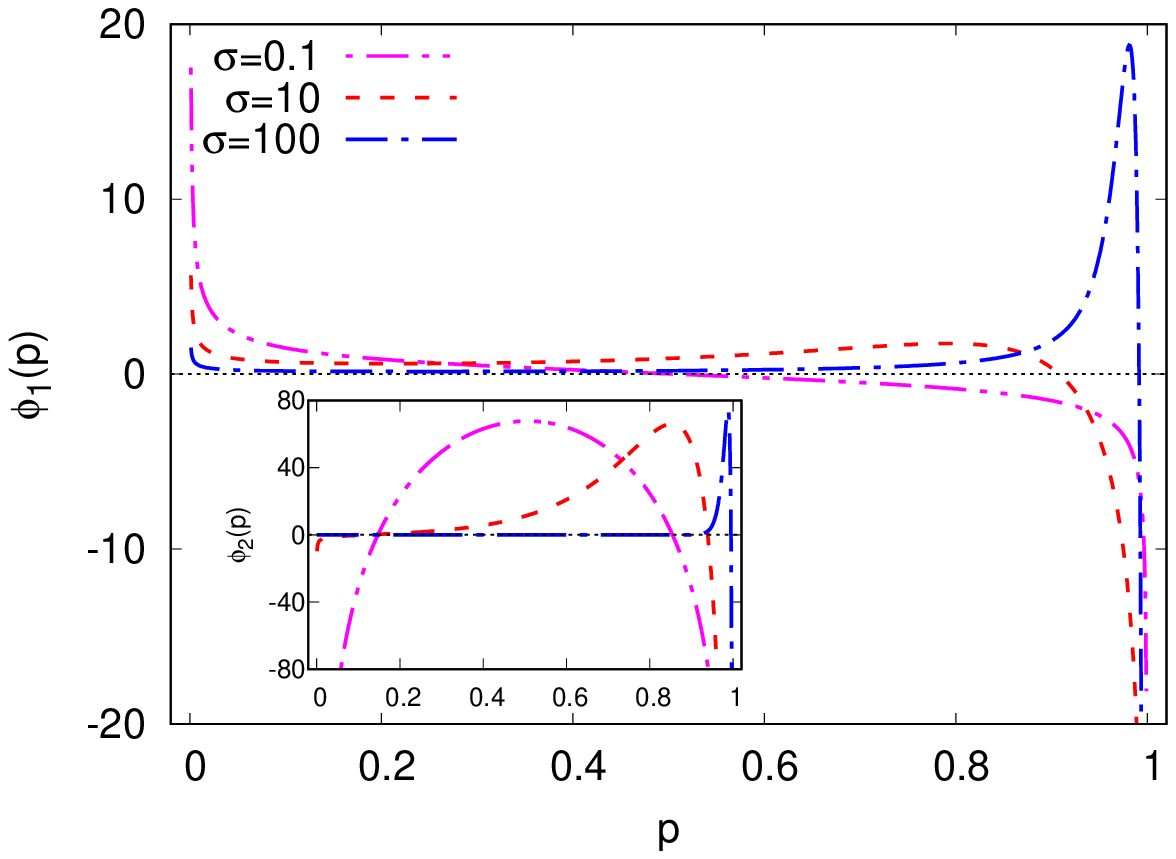}
 \includegraphics[width=0.8\textwidth]{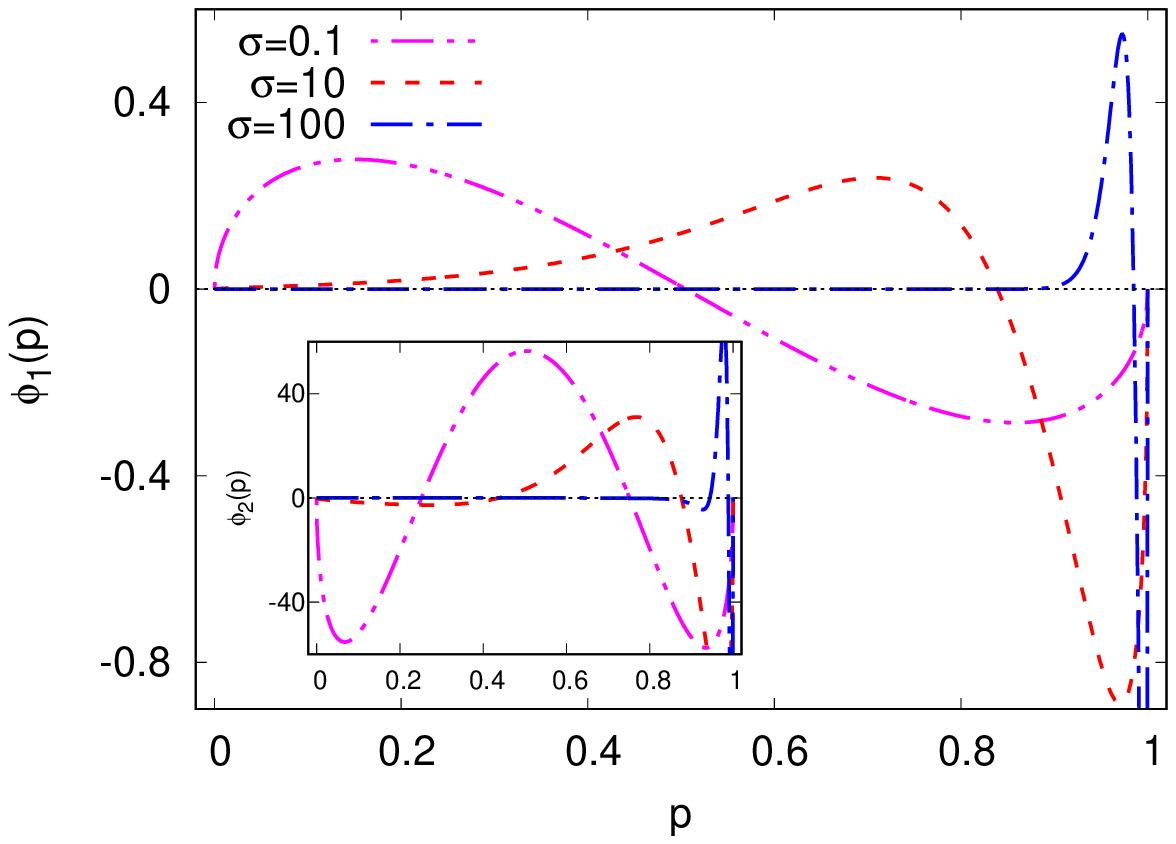}
\caption{First two excited states, $\phi_1(p)$ (main) and $\phi_2(p)$ (inset) obtained numerically using (\ref{jacobi}) and (\ref{cneqn}) for $\mu=1/2$ (top panel) and $3/2$ (bottom panel), and $K=1000$ for various selection strengths.}
\label{fig_eigenfn}
\end{figure}

Figure \ref{fig_eigenfn} shows the numerical results for the eigenfunctions $\phi_1(p)$ and $\phi_2(p)$ for various values of $\sigma$ and large $K$. As expected, the $\ell$th excited state has $\ell$ nodes whose location depends on $\mu$ and $\sigma$. Other than this feature, the distribution is qualitatively similar to that in the stationary state given by (\ref{equil}). When selection is absent, the eigenfunctions are symmetric about $p=1/2$ \cite{Kimura:1964}; for this reason, the first excited state for small $\sigma \lesssim 1$ has a node close to one half. But for stronger selection, the eigenstate is highly asymmetric and as in the stationary state, the excited states also peak close to the allele frequency equal to one.


\section{Weak selection limit}
\label{weak}

As the eigenvalues are known exactly for $\sigma=0$ \cite{Kimura:1964}, we can use a perturbation theory to determine $\lambda_\ell$'s for $\sigma \ll \mu$. On expanding the coefficient $c_n^{(\ell)}$ and eigenvalue $\lambda_\ell$ in a power series to quadratic orders in $\sigma$ and substituting them in (\ref{cneqn}), we find that the zeroth order term in $\sigma$ yields $\lambda_\ell(\sigma=0)=\ell (2 \mu+ \ell-1),  \ell=0, 1,...$ \cite{Kimura:1964}. The first order correction to the eigenvalue is found to be zero and the quadratic term in $\sigma$ leads to 
\be
\lambda_\ell (\sigma)\approx \lambda_\ell(\sigma=0)+ \frac{T_+(\ell+1) T_-(\ell+2)}{T_0(\ell+2)-T_0(\ell+1)}+\frac{T_+(\ell) T_-(\ell+1)}{T_0(\ell)-T_0(\ell+1)}.
\label{pertf}
\ee
From the above expression, we obtain the eigenvalue for the first excited state to be 
\be
\lambda_1 \approx 2\mu + \frac{\sigma^2 \mu}{(2\mu+2)(2\mu+3)}, 
\label{pert1}
\ee
which is in good agreement with the numerical data in Fig.~\ref{fig_lam1}. Equation (\ref{pertf}) also shows that as for $\sigma=0$, the gap between the consecutive eigenvalues increases with index $\ell$. 


\section{Strong selection limit}
\label{strong}

We now turn to the strong selection regime where $\sigma \gg \mu$. 
The coefficient $(T_0(n)+\lambda_\ell)/\sigma$ of $c_n^{(\ell)}$ in (\ref{cneqn}) shows that for $\sigma \to \infty$, a nontrivial eigenvalue for the excited states is obtained if $\lambda_\ell$ scales linearly with $\sigma$, in accordance with the numerical results shown in Fig.~\ref{fig_lam1}. One is then tempted to ignore the $T_0$ term altogether; however, as explained in Appendix~\ref{app_genfn}, this results in imaginary eigenvalues for any $K$. The correct limit procedure for strong selection is therefore to take $K \to \infty$ followed by $\sigma \to \infty$. In Sec.~\ref{scale} and Sec.~\ref{infsel}, we find the expansion coefficients for large and small $n$, respectively, for large, finite $\sigma$ when $K \to \infty$. Our analysis, however, does not yield eigenvalues that were studied in Sec.~\ref{num} numerically.  

\subsection{Scaling form for the expansion coefficients}
\label{scale}

Although we have not been able to obtain the eigenvalues analytically, a simple but accurate approximation for the coefficients $c_n^{(\ell)}, n \gg 1$ for large $\sigma$ can be obtained as follows. We again consider the coefficient $T_0(n)+\lambda_\ell$ of $c_n^{(\ell)}$: for large $\sigma$, $T_0 \sim n^2$ can be neglected in comparison to $\lambda_\ell \sim \sigma$ when $n \ll \sqrt{\sigma}$, and as a consequence, the expansion coefficient $c_n^{(\ell)}$ is independent of $\sigma$ for small $n$. But the eigenvalue can be ignored for $n \gg \sqrt{\sigma}$ and we may expect a $\sigma$-dependence for $c_n^{(\ell)}$ when $n$ is  large. These observations suggest that for $\sigma \gg 1$, the coefficient $|c_n^{(\ell)}|$ is of a scaling form, 
\be 
\frac{|c_n^{(\ell)}|}{|c_2^{(\ell)}|}=C_{\sigma,\mu} f_\ell \left(\frac{n}{\sqrt{\sigma}}\right), 
\label{sclfrm}
\ee
where the $\sigma$-dependence of $C$ is fixed using that the $|c_n^{(\ell)}|$ must be independent of $\sigma$ for small $n$.  

Since the $c_n^{(\ell)}$'s have alternating signs, (\ref{cneqn}) gives
 \be
 T_-(n) |c_{n-1}^{(\ell)}| +  T_{+}(n) |c_{n+1}^{(\ell)}| =  (T_0(n)+\lambda_\ell)|c_n^{(\ell)}| ~,~n \geq 2.
\label{modcneqn}
\ee
Using the scaling form (\ref{sclfrm}) in the above equation and collecting terms to order $\sigma$, we arrive at a {\it first order} linear differential equation for the scaling function $f_\ell(u)$ given by (see also Appendix~\ref{app_genfn})
\be
\frac{u}{2} f_\ell'(u)+\left(u^2-\mu-\kappa_\ell+\frac{1}{2} \right) f_\ell(u) =0, 
\label{scldiff}
\ee
where $\kappa_\ell=\lim_{\sigma \to \infty} {\lambda_\ell}/{\sigma}$. On solving the above differential equation, we  find that
\be
|c_n^{(\ell)}| \approx C_\mu n^{2 \mu+2 \kappa_\ell-1} e^{-{n^2}/{\sigma}}.
\label{sclfn}
\ee
In Fig.~\ref{fig_scal}, the numerically obtained expansion coefficients $|c_n^{(1)}|$ and the above expression along with the conjectured eigenvalue (\ref{firstlambda}) for the first excited state are compared and we find a very good agreement, except for small $n$ where the scaling limit is not valid. Equation (\ref{scldiff}) also shows that the scaling function $f_\ell(u)$ has a turning point at $u^*_\ell=\sqrt{\mu+\kappa_\ell-\frac{1}{2}}$; for the first excited state, using the conjecture (\ref{firstlambda}), we find that $|c_n^{(1)}|$ is a nonmonotonic function for $\mu > 1/4$ and decreases monotonically otherwise. We have numerically verified that (\ref{sclfn}) works well for higher excited states also (data not shown). 

\begin{figure}
\includegraphics[width=0.8\textwidth]{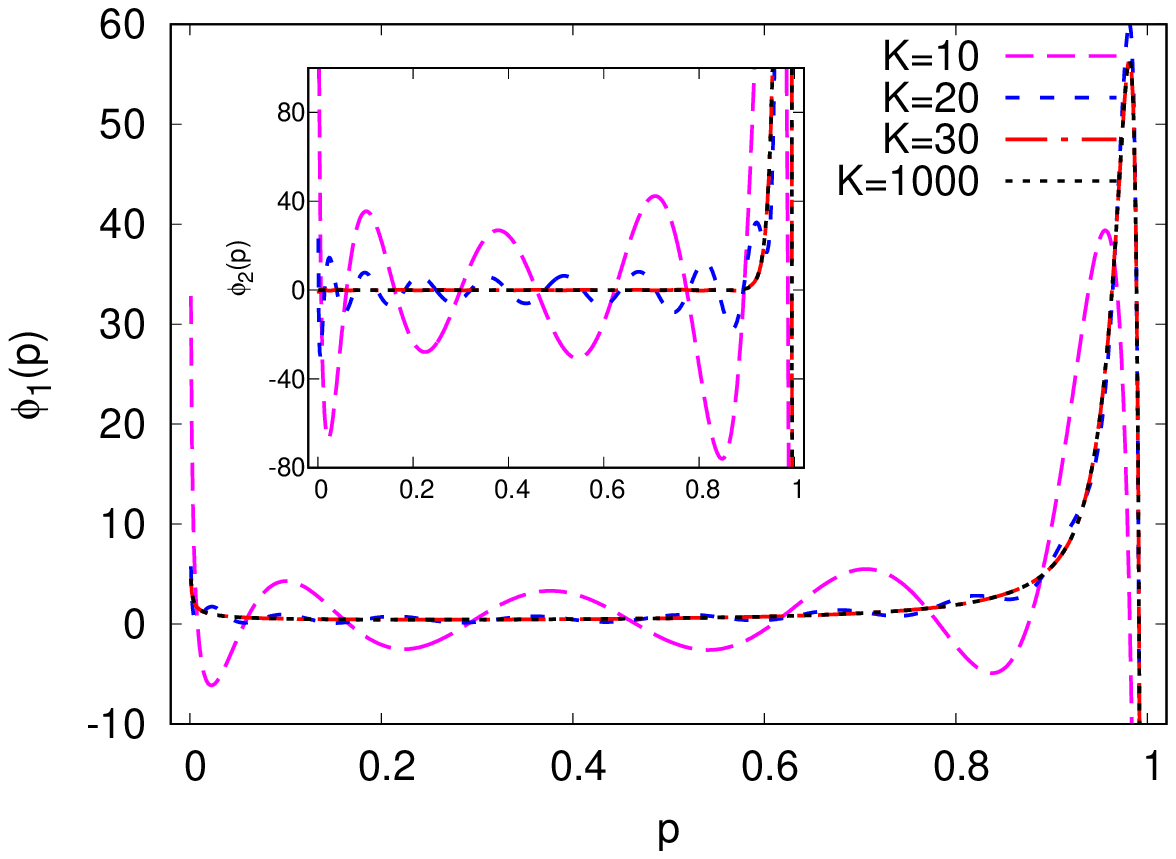}
\includegraphics[width=0.8\textwidth]{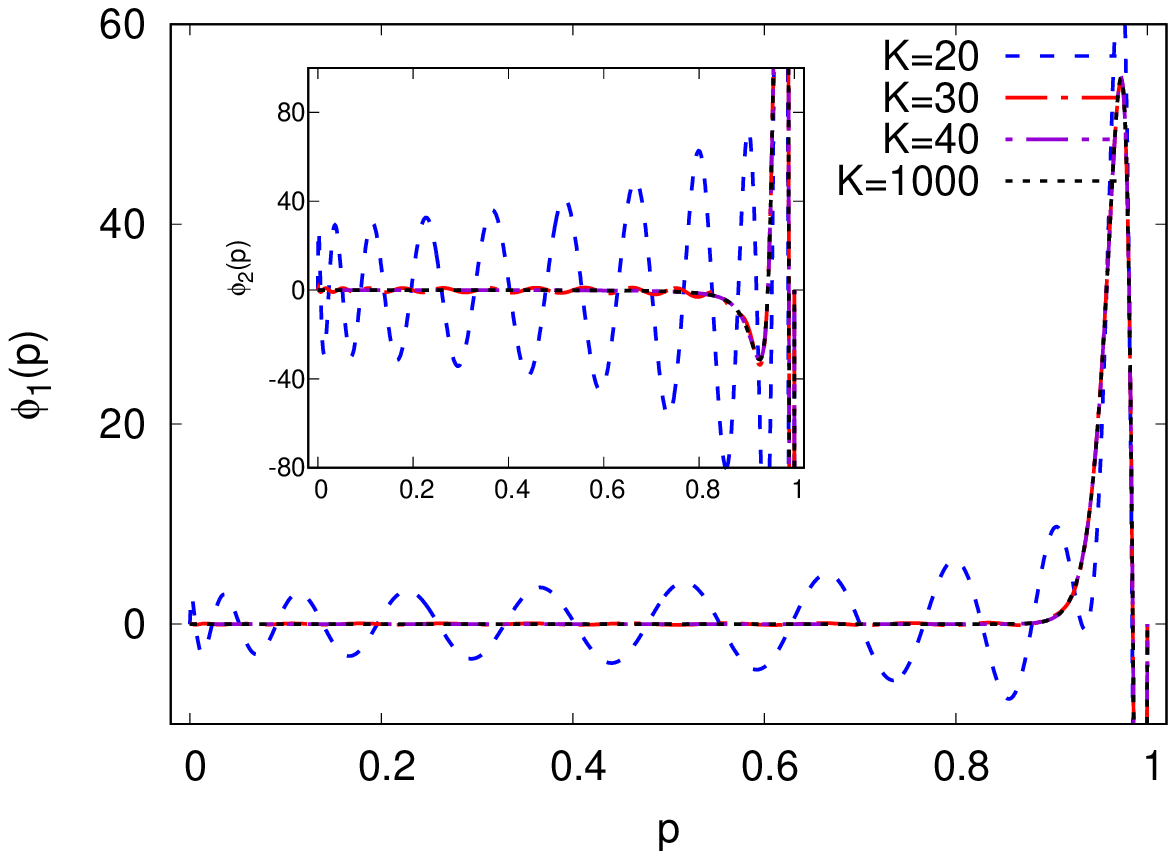}
\caption{First two excited states, $\phi_1(p)$ (main) and $\phi_2(p)$ (inset) obtained numerically when the orthogonal expansion series (\ref{jacobi}) is terminated at $n=K+1$ for strong selection ($\sigma=100$) and mutation rate, $\mu=1/2$ (top panel) and $3/2$ (bottom panel).}
\label{fig_Kdepn}
\end{figure}

Equation (\ref{sclfn}) suggests that as the expansion coefficients decay fast for large $n$, 
it may be sufficient to keep terms up to $n \lesssim \sqrt{\sigma}$ for the evaluation of sum (\ref{jacobi}) for the eigenfunction. This expectation is tested in Fig.~\ref{fig_Kdepn} for two eigenfunctions, and we find that when the sum is terminated at small $K \sim \sqrt{\sigma}$, the result matches well with those obtained with $K \gg \sigma$. We remark that at small $K$, the eigenfunction is seen to have several nodes. But, as already shown in Fig.~\ref{fig_eigenfn}, the eigenfunction $\phi_\ell(p)$ has $\ell$ nodes for large enough $K$. 

\subsection{Expansion coefficients for infinite selection}
\label{infsel}

\begin{figure}
\includegraphics[width=0.8\textwidth]{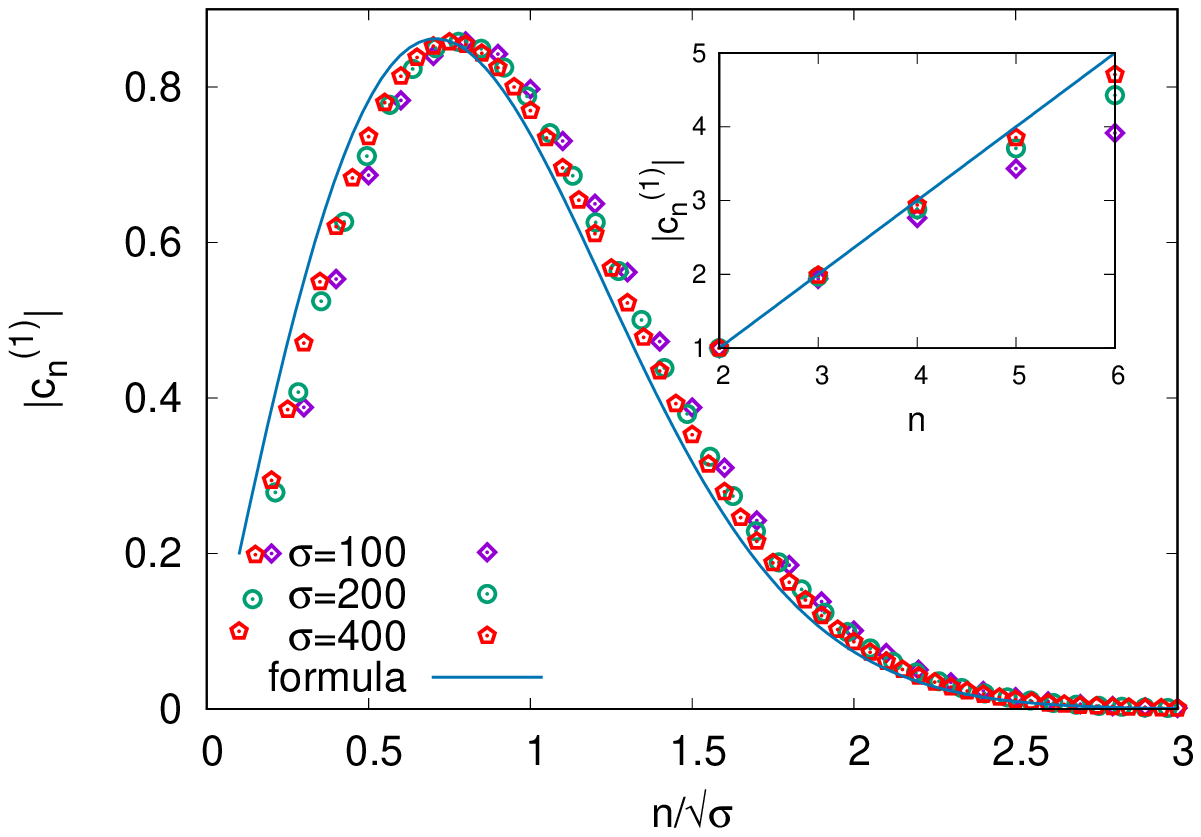}
\includegraphics[width=0.8\textwidth]{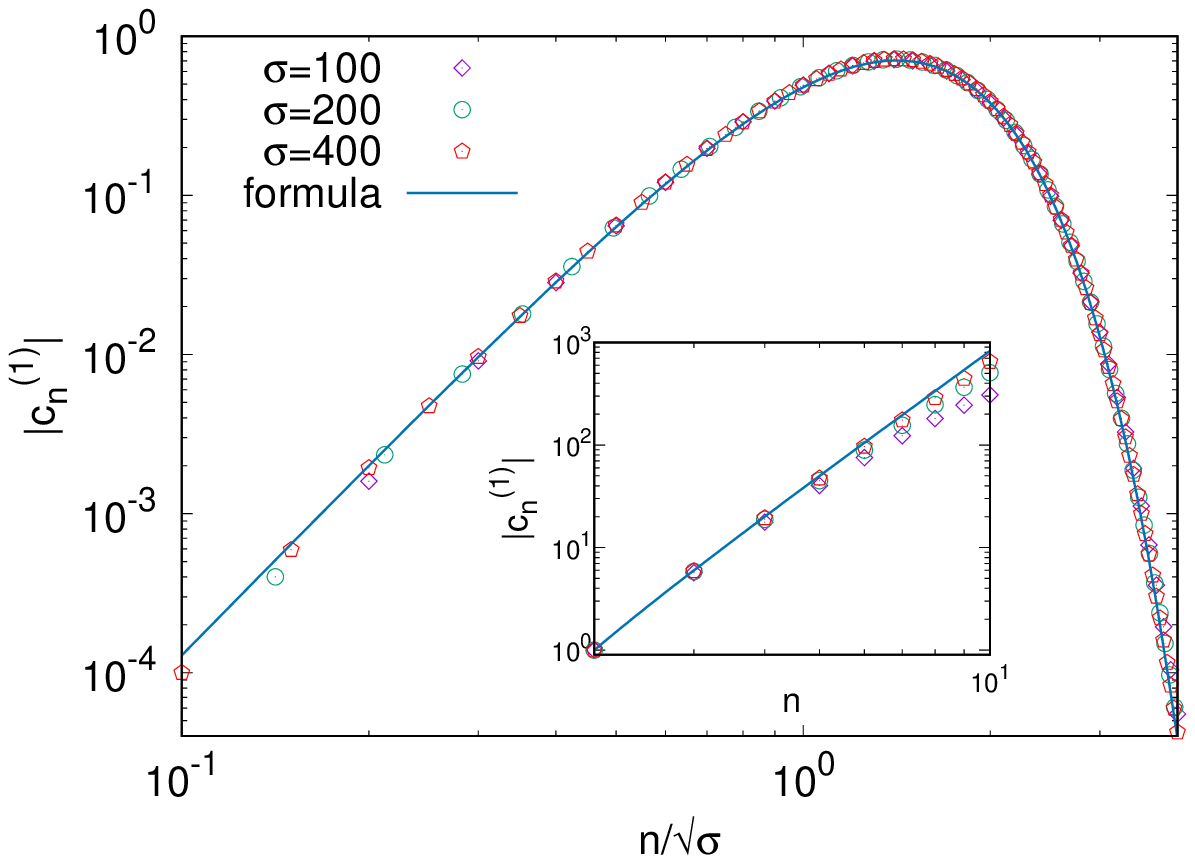}
\caption{Main: Expansion coefficient $|c_n^{(1)}|$ obtained numerically (points) for various $\sigma$ and analytical expression (\ref{sclfn}) in scaling limit for $\mu=1/2$ (top) and $3/2$ (bottom), and $K=1000$. The inset shows the expansion coefficient for small $n$ and is compared with (\ref{csoln}).}
 \label{fig_scal}
\end{figure}

As shown in Fig.~\ref{fig_scal}, the expansion coefficients in (\ref{sclfn}) do not match with the numerical results for small $n$. Here, we show that these can be found exactly for special values of  $\mu$. For this purpose, it is useful to define 
\be
d_n^{(\ell)}=\bigg \rvert\frac{c_n^{(\ell)}}{6-4 \mu-4 n}\bigg \rvert~,~n \geq 2.
\label{dndefn} 
\ee
As shown in Appendix~\ref{app_genfn}, the generating function 
${\cal D}_\ell(z)=\sum_{n=2}^\infty d_n^{(\ell)} z^n$ obeys a third order differential equation given by (\ref{app_order3}); however, for $\sigma \to \infty$, we obtain
\be
{\cal D}_\ell(z)=
\begin{cases}
\frac{z}{4 \kappa_\ell} \left(\left(\frac{1+z}{1-z} \right)^{2 \kappa_\ell}-1 \right) &~~ (\mu=1/2) \\
\left(\frac{z}{1-z^2} \right)^2 \left(\frac{1+z}{1-z} \right)^{2 \kappa_\ell} &~~(\mu=3/2)
\end{cases}
\label{Dsoln}
\ee
on choosing $d_2^{(\ell)} =1$. Using the conjecture (\ref{firstlambda}), we get
\be
\frac{|c_n^{(1)}|}{|c_2^{(1)}|} \stackrel{\sigma \to \infty}{\longrightarrow}
\begin{cases}
n-1&~~ (\mu=1/2) \\
\frac{n^2 (n^2-1)}{12} &~~(\mu=3/2)
\end{cases},
\label{csoln}
\ee
which are consistent with the power law scaling in (\ref{sclfn}) and the numerical data in the inset of Fig.~\ref{fig_scal} for small $n$. 

\section{Discussion}

While the confluent Heun equation frequently appears in various problems in physics \cite{Hortacsu:2018}, here, for the first time, we have made a connection between this equation and a standard population-genetics model defined by (\ref{ufullFP}). This relationship is useful, especially for biologists \cite{Wang:2004,Bollback:2008,Gutenkunst:2009}, as one can simply use the standard packages (such as Maple and latest version of Mathematica) to solve the partial differential equation (\ref{ufullFP}) numerically. However, in \cite{Song:2012} and here, an orthogonal series expansion is used to recast the problem as an eigenvalue problem (\ref{cneqn}) which can also be easily implemented numerically and is, perhaps, more amenable to analysis (see, for e.g., \cite{Leaver:1985}). 

Here we have studied the eigenvalues and eigenfunctions of the confluent Heun operator in some detail. Our main result for the first eigenvalue $\lambda_1$ which is inversely proportional to the relaxation time is summarized in Fig.~\ref{fig_lam1}.  For strong selection ($\sigma \gg \mu$), we find that there is a transition at mutation rate $\mu=1$ and the eigenvalue $\lambda_1$ is independent of mutation rate for $1 < \mu \ll \sigma$. In contrast, in the absence of selection, the relaxation time decreases monotonically as $1/\mu$ \cite{Kimura:1964}.

Although we have produced strong numerical evidence for the behavior of the eigenvalues with model parameters, for the reasons described in Sec.~\ref{intro}, it seems very difficult to make analytical progress. However, in the limit of strong selection, an analytical understanding of (\ref{firstlambda}), perhaps using a WKB approximation \cite{Ferrari:1984}, may be possible.  

\clearpage


\appendix
\makeatletter
\renewcommand{\@seccntformat}[1]{Appendix \csname the#1\endcsname\quad}
\makeatother
\renewcommand{\thesection}{\Alph{section}}
\numberwithin{equation}{section}



\section{Frobenius series expansion}
\label{app_frob}

Since $p=0$ is a regular singular point of the confluent Heun equation, we can expand the eigenfunction in a Frobenius series by writing $\phi_\ell(p)=p^{a} \sum_{n=0}^\infty f_n^{(\ell)} p^n, |p| < 1$ \cite{Mathews:1970}. Substituting it in (\ref{heun}) and setting the coefficient of $p^{a-1}$ to zero, we find the indicial exponents to be $a=0$ and $\mu-1$. For $\mu \neq 1$, the eigenfunction can be written as
\be
\phi_\ell(p)={\tilde a}_1 H_C(-\sigma,1-\mu,1-\mu,-\mu \sigma,\eta_\ell,p)+ {\tilde a}_2 p^{\mu-1} H_C(-\sigma,\mu-1,1-\mu,-\mu \sigma,\eta_\ell,p), 
\ee
where $\eta_\ell=(1-2 \lambda_\ell +\mu (\sigma-\mu))/2$ and $H_C(\alpha, \beta, \gamma, \delta, \eta, p)$ is the confluent Heun function  \cite{Fiziev:2010}. For $\mu=1$, the first solution has a logarithmic singularity at $p=0$. In either case, the vanishing current boundary condition (\ref{bdry}) at $p=0$ yields ${\tilde a}_1=0$. 

The coefficients of terms of ${\cal O}(p^a)$ or higher in the Frobenius series lead to a 3-term recursion relation for $f_n^{(\ell)}$'s given by
\be
(n+\mu) (n+1) f_{n+1}^{(\ell)} + [(n+\mu-1) (\mu-\sigma-n-2)+\nu_\ell] f_n^{(\ell)} +\sigma  (n+\mu) f_{n-1}^{(\ell)}=0~,~n=0, 1,...
\ee
with $f_{-1}^{(\ell)} =0$. The reflecting boundary condition at $p=1$ imposes the condition $\sum_{n=0}^\infty f_n^{(\ell)}=0$. Note that unlike in the expansion (\ref{cneqn}), here the parameter $\sigma$ appears in the coefficient of $f_n^{(\ell)}$ and $f_{n-1}^{(\ell)}$. 

\section{Orthogonal polynomial expansion}
\label{app_rec}

We begin with the observation that in the absence of selection, the eigenfunctions are exactly given by \cite{Kimura:1964}
\be
\phi_\ell(p,\sigma=0) \propto (p q)^{\mu-1} P_\ell^{(\mu-1,\mu-1)}(1-2 p), 
\ee
with eigenvalue $\lambda_\ell(\sigma=0)=\ell (2 \mu+\ell-1), \ell=0, 1, ..$. For nonzero selection, we therefore write
\be
\phi_\ell(p,\sigma)=(p q)^{\mu-1} \sum_{n=0}^\infty a_n^{(\ell)} P_n^{(\mu-1,\mu-1)}(1-2 p). 
\label{app_expn}
\ee
As this expansion is valid for $|p| < 1$, in order to impose the boundary condition (\ref{bdry}) at $p=1$, we used the relationship between Jacobi polynomials and Gauss hypergeometric function and the connection formulae for the latter \cite{Abramowitz:1964},  and verified that (\ref{bdry}) is indeed satisfied. 

Substituting the expansion (\ref{app_expn}) in the confluent Heun equation (\ref{heun}) and using the expression for the second derivative of Jacobi polynomial (see 22.6.2 of \cite{Abramowitz:1964}), we obtain
\be
-\sigma (p q \phi_\ell'(p)-2 p \phi_\ell)= \sum_{n=0}^\infty a_n^{(\ell)} [n (2 \mu-1+n)-{\lambda_\ell}+{\sigma} ] (1-x^2)^{\mu-1} P_n^{(\mu-1,\mu-1)}(x),
\label{app_simple1}
\ee
where $x=1-2 p$. It can be verified that the above equation reproduces the eigenspectrum in the absence of selection. 

To proceed further, we need the following identities:\\
1. From 22.8.1 of \cite{Abramowitz:1964}, we get 
\be
(1-x^2) \frac{d P_n^{(\mu-1,\mu-1)}(x)}{dx}=-n x P_n^{(\mu-1,\mu-1)}(x)+(n+\mu-1)P_{n-1}^{(\mu-1,\mu-1)}(x). 
\label{app_id2}
\ee
2.  Furthermore, 22.7.15 and 22.7.18 of \cite{Abramowitz:1964}, gives
\bea
(1-x) P_n^{(\mu-1,\mu-1)}(x)&=&P_n^{(\mu-1,\mu-1)}(x) -\frac{n+\mu-1}{2 \mu + 2 n-1} P_{n-1}^{(\mu-1,\mu-1)}(x) \nonumber \\
&-&\frac{(n+1)(2 \mu + n-1)}{(\mu + n) (2 \mu + 2 n-1)}P_{n+1}^{(\mu-1,\mu-1)}(x). 
\label{app_id1}
\eea
Using (\ref{app_id2}) and (\ref{app_id1}) in (\ref{app_simple1}), we find that 
\bea
&&\frac{p q \phi_\ell'(p)-2 p \phi_\ell(p)}{(1-x^2)^{\mu-1}} +\sum_{n=0}^\infty  a_n^{(\ell)} P_n^{(\mu-1,\mu-1)}(x) \nonumber \\
&=& \sum_{n=1}^\infty  \left[ 
 \frac{n (2 \mu + n-2) (2 \mu + n-1) }{2 (\mu + n-1) (2 \mu + 2 n-3)}a_{n-1}^{(\ell)}-\frac{(\mu + n) n }{2 (1 + 2 \mu + 2 n)} a_{n+1}^{(\ell)} \right] P_n^{(\mu-1,\mu-1)}(x). \nonumber
\eea
On matching the coefficient of $P_n^{(\mu-1,\mu-1)}(x)$ on both sides of the equation, we finally obtain (\ref{jacobi})-(\ref{cneqn}).

\section{Stationary state}
\label{app_ss}

For ${\lambda}_0=0$, we can find the coefficient $c_n^{(0)}$ using the known steady state distribution (\ref{equil}) and orthonormality property of Jacobi polynomials. From (\ref{jacobi}), we have 
\be
\frac{e^{\sigma p}}{Z}=\sum_{m=0}^\infty c_{m+1}^{(0)} \frac{\Gamma(m+1)}{\Gamma(m+\mu)} P_{m}^{(\mu-1,\mu-1)}(1-2 p), 
\ee
where $Z$ is the normalization constant. Using the orthonormality property (\ref{jacorth}) of Jacobi polynomials, we arrive at
\bea
c_{m+1}^{(0)} \frac{\Gamma(m+1)}{\Gamma(m+\mu)} h_m Z &=& \int_{-1}^1 dx (1-x^2)^{\mu-1} P_m^{(\mu-1,\mu-1)}(x) e^{\frac{\sigma (1-x)}{2}} \\
&=& \frac{\sqrt{\pi} \Gamma(m+\mu) e^{\frac{\sigma}{2}} }{m ! \Gamma(m+\mu+\frac{1}{2})} \left(\frac{-\sigma}{4} \right)^m {_1}F_1(1,m+\mu+\frac{1}{2},\frac{\sigma^2}{16})
\eea
where ${_1}F_1(1;b;x)$ is confluent hypergeometric function \cite{Abramowitz:1964}. 
It is straightforward to check that the expansion choice of \cite{Song:2012} leads to essentially the same result as above. 


\section{Generating function for strong selection}
\label{app_genfn}

The expansion coefficient $d_n$ defined in (\ref{dndefn}) obeys the following recursion equation,
\be
(2 \mu+n-2) (2 \mu+n-3) d_{n-1}^{(\ell)} + n (1-n) d_{n+1}^{(\ell)} = \left[\kappa_\ell+\frac{(1-n)(2 \mu+n-2)}{\sigma}\right] (6-4\mu-4n) d_n^{(\ell)}. 
\label{app_dneqn}
\ee
The above form allows one to write a differential equation for the generating function ${\cal D}(z)=\sum_{n=2}^\infty d_n z^n$ that  obeys the following third order ordinary differential equation,  
\bea
\hspace{-1in}z (z^2-1) {\cal D}'' + 2 (1+(2 \mu-1)z^2) {\cal D}' +2 (z (1-\mu)(2 \mu-1)-z^{-1}){\cal D} -\kappa [(6-4 \mu) {\cal D}-4 z {\cal D}' ]\nonumber \\
\hspace{-1in}=\frac{1}{\sigma} \left[ 4 z^3 {\cal D}'''+ 6 z^2 (2 \mu-1) {\cal D}''+4 z (\mu-1) (2 \mu-3) {\cal D}' +4 (1-\mu) (2 \mu-3) {\cal D}\right],
\label{app_order3}
\eea
where we have dropped the eigenvalue label for brevity. 
The above equation does not appear to be solvable; however, in the scaling limit, $z \to 1, \sigma \to \infty$ such that $x=(1-z) \sqrt{\sigma}$ is finite, we obtain
\be
\frac{d^3 {\cal D}}{d x^3}-\frac{x}{2} \frac{d^2 {\cal D}}{d x^2}- (\eta+\mu) \frac{d {\cal D}}{d x}=0, 
\ee
that yields
\be
{\cal D}(x)={\tilde c}_1 \frac{H_{1-2 \eta-2 \mu}(\frac{x}{2})}{1-2 \eta-2 \mu}+{\tilde c}_2 x {_1}F_1\left(\eta+\mu,\frac{3}{2},\frac{x^2}{4} \right)+{\tilde c}_3, 
\ee
where $H_n(x)$ is the Hermite function and ${_1}F_1(a,b,z)$ is the Kummer confluent hypergeometric function \cite{Abramowitz:1964}. As $n \geq 2$, the constant ${\tilde c}_3=0$; furthermore, we numerically found that the inverse Laplace transform of the second term on the RHS grows exponentially with $n$ and therefore ${\tilde c}_2=0$. The asymptotic expansion of the Hermite function then yields (\ref{sclfn}). 

For $\mu=1/2$ and $3/2$, the recursion equation (\ref{app_dneqn}) for $d_n, n \geq 2$ simplifies and leads to following second order differential equation for ${\cal D}$:
\bea
(1-z^2) {\cal D}'+(z-z^{-1}){\cal D}-d_2 z &=& 4 [\kappa {\cal D}- \frac{1}{\sigma} (z^2 {\cal D}''-z {\cal D}'+{\cal D})] ~~(\mu=1/2)
\label{app_1h} \nonumber \\
(1-z^2) {\cal D}'-2 (z+z^{-1}){\cal D} &=& 4 [\kappa {\cal D}- \frac{1}{\sigma} (z^2 {\cal D}''+z {\cal D}'-{\cal D})]  ~~(\mu=3/2) 
 \label{app_3h}. \nonumber
\eea
Both of these equations have an irregular singularity at $z=0$ and infinity, each of rank $1$ which is the same as doubly-confluent Heun equation \cite{heunsite,Hortacsu:2018}. 
Thus the generating function can not be reduced to simpler functions. However, for $\sigma \to \infty$, we obtain (\ref{Dsoln}) in the main text. 

We note that the matrix ${\bf T}$ in (\ref{cneqn}) is not normal (that is, it does not commute with its transpose), and therefore there is no guarantee that its eigenvalues would be real \cite{Mathews:1970}. For finite $\sigma$, we found numerically that the first few eigenvalues are real and the rest are complex for finite $K$ but the imaginary part of the eigenvalue decreases towards zero with increasing $K$. For infinite $\sigma$ and finite $K$, the generating function ${\cal D}(z)=\sum_{n=2}^K z^n d_n$ for $\mu=1/2$  obeys the inhomogeneous equation, 
\be
(1-z^2) {\cal D}'+(z-z^{-1}){\cal D}-d_2 z= 4 \kappa {\cal D}-(K-1) z^{K+1} d_K. 
\ee
On demanding that the solution of the above equation does not have terms of order $z^{K+1}$ and higher, we find that one of the eigenvalues is zero (if $K$ is odd) and the rest are complex for any $K$. This discussion thus reiterates the point that the eigenvalues of interest are obtained if the strong selection limit is taken after $K \to \infty$.


\clearpage

\clearpage

\clearpage

\clearpage

\clearpage

\clearpage
\makeatletter
\renewcommand\@biblabel[1]{[#1]}
\makeatother
\providecommand{\newblock}{}


\end{document}